\documentstyle[aps,prb,preprint,floats,tighten,epsfig]{revtex}

\begin{document}
\thispagestyle{empty}

\title{Recent theories of glasses as out of equilibrium systems
\footnote{To be published in the special issue  `Physics of Glasses'
  of the Comptes Rendus de Physique de
  l'Acad\'emie des Sciences.}}

\author{Jorge Kurchan}
\address{ 
\it P.M.M.H. Ecole Sup{\'e}rieure de Physique et Chimie Industrielles,
\\
10, rue Vauquelin, 75231 Paris CEDEX 05,  France}

\date\today

\maketitle


\begin{abstract}
 We discuss a theoretical  approach  to  structural glasses  
 as they happen in real situations.
Older ideas based on  `configurational entropy', on `fictive
temperatures' and on Edwards' `compactivity' are sharpened and unified
in an out of equilibrium context.
The picture is such that it may be supported or disproved by 
an experimental test, which we describe.
\end{abstract}
\vspace{.5cm}
 

 In this note I briefly review  a rather general 
 picture of structural glasses that has emerged in the last dozen years or so,
 and that  to my knowledge has not been summarised elsewhere from
 quite this point of view.
 To the extent that this scenario applies, one can make a theory of
 glasses {\em as they happen}, any reference to ideal infinitely  slow
 annealings being superfluous.
 
 Two sets of older ideas ---  one based on configurational entropy
 put forward by Kauzmann, Adam, Gibbs and Di Marzio \cite{Kauzmann}
  and the other on aging, `fictive temperatures' \cite{Tool},
 and Edwards' `compactivity' \cite{Sam}  --- are unified  and
 developed in two ways:

\begin{itemize}

\item They either admit a quantitative definition, or, in certain aspects,
  their inherent ambiguity is uncovered.

\item  Models and approximations are identified   
 within which one can calculate everything analytically, and the
 scenario holds strictly.     
 This opens the way for systematic improvements
 of these schemes, a line that is currently pursued.
 Solvable cases have indicated  
 that certain features of the older ideas  had to be reconsidered, of
which we shall see some instances below.

\end{itemize}
\vspace{1cm}

{\bf 1. Aging}

\vspace{1cm}

 Consider a supercooled liquid. In order to probe its dynamics we
 shall use an autocorrelation function, say, 
 $C(t_1-t_2))= <\rho_k^*(t_1)\rho_k(t_2)>$. In Fig. 1 we show how this looks
 like when plotted in terms of $\log(t)=\log(t_1-t_2)$ for two
 different temperatures $T_a>T_b$.
 There is a fast relaxation which corresponds to rapid motion of the
 particles `inside a cage' formed by its neighbours,
 much as atoms move around their lattice position  in  a crystal.  
 On much longer timescales, the actual `cages' will rearrange:
 these are the `structural' or `alpha' relaxations.
 While the cage motion is weakly dependent on the temperature, the 
 structural motion slows dramatically as we cool the system.

 In Fig. 1 we also show two arbitrary ways of defining an
 alpha-relaxation time $t_\alpha(T)$.  If, as we decrease $T$, we find
 that $t_\alpha'(T)$ and $t_\alpha(T)$ stay of the same order,
 then we have only one slow timescale, i.e. we have in all a two-step process.
 For the moment it seems that this is the situation for structural
 glasses.
 Another possibility, encountered in mean-field {\em spin} glasses, is
 that on tuning the appropriate parameter any two timescales go to
 infinity at different paces, the ratio $t_\alpha/t_{\alpha}'$
 becoming larger and larger: this is a multi-timescale situation.
 This serves to underline the fact that when one talks
 about several timescales, the concept becomes clear-cut only in an
 appropriate limit in which they separate completely. 
 This will turn out to be relevant in what follows. 

\begin{figure}
\centerline{\epsfxsize=9cm
\epsffile{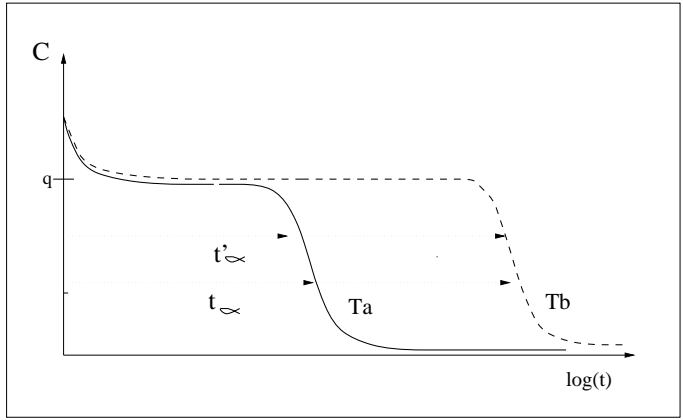}
}
\caption{A correlation.}
\label{coins}
\end{figure}

 An alternative way to describe the relaxation is by considering  
 the {\em integrated} response $\chi(t_1-t_2)$
 conjugated to $C$: in this case   $\chi(t_1-t_2)$ is
 the compressibility of the mode $k$ with respect to a field acting
 continuously from $t_2$ to $t_1$.   
 In equilibrium we have that $T \chi(t_1-t_2)= C(t_1-t_1)-C(t_1-t_2)$,
 by virtue of the fluctuation-dissipation theorem (FDT), see Fig. 2.

\begin{figure}
\centerline{\epsfxsize=9cm
\epsffile{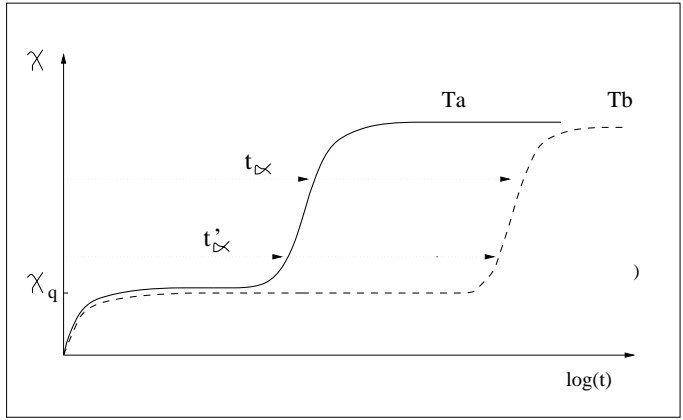}
}
\caption{The associated response.}
\label{coinsa}
\end{figure}

 In cases in which we have a two-step relaxation, we can describe the
 slow dynamics of the system by a plot as in Fig. 3.
 Such plots are in general done with the viscosity instead of
 $t_\alpha$ in the $y$-axis.

 At this point the tradition is to embark in a detailed
 discussion of the form of the curve Fig. 3. Is there a true
 divergence at some $T_o>0$ or not? If so, what is the nature of
 the equilibrium thermodynamic state below $T_o$?
 Here I take the point of view that these questions, although of 
 obvious interest, are not  the most urgent as far as understanding 
 glasses in real life, and one can make progress without addressing them.
 Let us here content ourselves with mentioning that there are some
 glasses (like plastics) that may have a $T_o>0$, while others are 
 expected to have a finite $t_\alpha$ at all temperatures above zero.
 Note that glasses from this last group, of which `window' glasses are 
 likely to be 
an example, are described from the equilibrium thermodynamic  point 
 of view as being at all temperatures just  supercooled  liquids.

\begin{figure}
\centerline{\epsfxsize=9cm
\epsffile{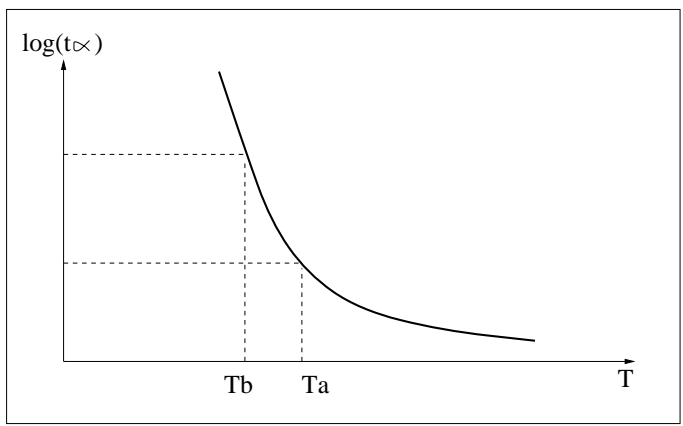}
}
\caption{Structural timescale versus temperature.}
\label{coinsb}
\end{figure}

Consider now a system having at $T_a$ a timescale $t_\alpha$ of ten
minutes, and at $T_b$ of an hour. What will happen if, from an equilibrium
situation at $T_a$   we lower the bath's temperature to  $T_b$
{\em in one minute}?
In the instant $t_w=0$ just after the quench,
while the  rattling of particles inside their cages will
quickly adapt to $T_b$, the cages themselves will not have the time 
to adapt to their new situation, as this needs structural
rearrangements taking times $t_w$ much larger than an  hour.
 
If we measure the correlations starting from
different $t_w$, it is quite clear that 
we will observe a relaxation as in Fig. 4.
The $\alpha$-relaxation time grows with $t_w$ and eventually saturates at
the new value --- and this can also be said about the viscosity. 
In other words, the two-time correlation function becomes waiting-time
 dependent: $C=C(t+t_w,t_w)$.
In particular, if there is a true divergence in the equilibrium value
of $t_\alpha(T)$
at some $T_o>T_b$, the $t_w$-dependence will stay forever.
{\em This is the aging phenomenon}:  quantities depend on the
waiting time for a long (eventually infinite) time $t_w >> t_\alpha(T_b)$.
The situation described above will always be met in a real-life cooling
procedure, as sooner or later the cooling time will be too short
compared with the equilibrium $t_\alpha(T)$.
Our problem is to understand the $t_w<<t_\alpha(T)$ regime that ensues,
 whose phenomenology, we argue, does not depend
crucially on whether $t_\alpha(T)$ is actually infinite or not.

\begin{figure}
\centerline{\epsfxsize=9cm
\epsffile{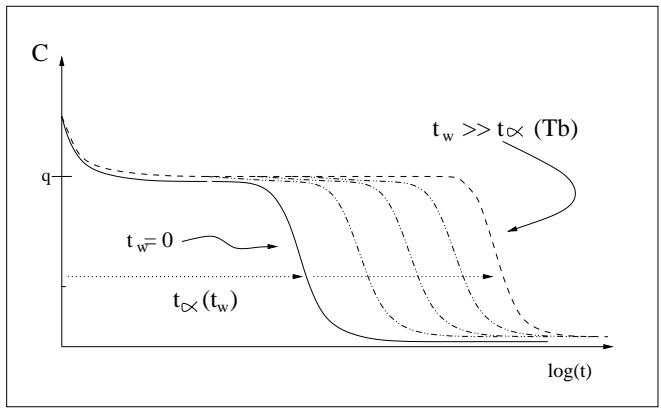}
}
\caption{Aging in correlations.}
\label{coinsc}
\end{figure}
 
 If instead of a correlation function we look at the associated
 response, then we also see aging.  Figure  5 is a sketch of how
 this would look. If we are considering a  uniform compression ($k=0$),
 the vertical arrow shows the beginning of the
 `creep' deformation,  as studied in
 the seminal experiments by Struik \cite{Struik}. 

\begin{figure}
\centerline{\epsfxsize=9cm
\epsffile{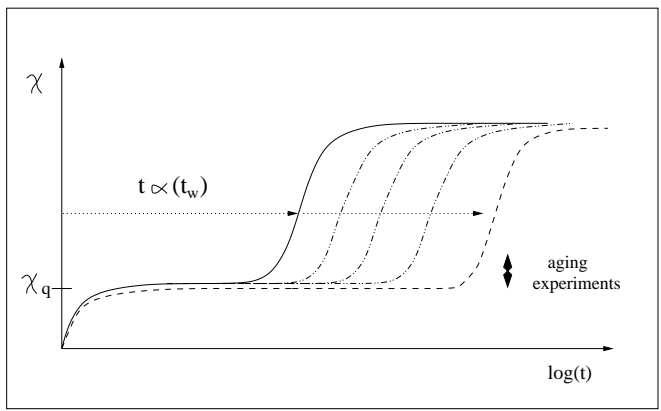}
}
\caption{Aging in responses.}
\label{coinsd}
\end{figure}

We have mentioned that when the system is quenched
it rapidly adapts its fast, rattling motion to the new temperature.
Its structural motion instead remembers the situation in which it lost pace
and  fell out of equilibrium: thus the idea of using the 
 temperature
 at which this happened as an extra  history-encoding state 
variable --- the `fictive temperature'  \cite{Tool}. 
This old idea, which had been somewhat abandoned by theorists~\cite{Samf}
 finds a place and a
 clear definition within the  framework of this and the following section.

 Consider a  plot of the $y$-axis of Fig. 1 versus the $y$
 axis of Fig. 2, where time acts as a parameter.
 The fluctuation-dissipation theorem tells us that for temperatures
 $T_a$ and  $T_b$ we obtain straight lines  with 
gradients 
 $-1/T_a$ and $-1/T_b$, respectively.
 What happens if we attempt to put together in the same way the aging
 curves Figs. 4 and 5, using {\em both} $t$ and $t_w$ as parameters? 
 A sketch of this is shown in Fig. 6.
 Soon after the quench, all curves superpose in the range
 $(C>q,\chi<\chi_o)$ on a straight line with gradient  $-1/T_b$: just
 as we would have expected to be the case in a system thermalised to
 the new temperature $T_b$. This is not unreasonable, as it
 corresponds to the times `inside a cage' (cfr. Figs. 4 and 5).
 The surprising result appears in the slow relaxation range 
$(C<q,\chi>\chi_q)$ : all the  curves superpose on {\em another
 straight line} with gradient (say)  $-1/T_{eff}$ (curve (1) of
 Fig. 6).
Then, on a much slower timescale (many $\alpha$-relaxations), the straight
line slowly drifts (curve (2) in Fig.6) eventually working its way 
to the dashed line of gradient $-1/T_b$, at which time the system has
 finally re-thermalised and aging has finished.

\begin{figure}
\centerline{\epsfxsize=9cm
\epsffile{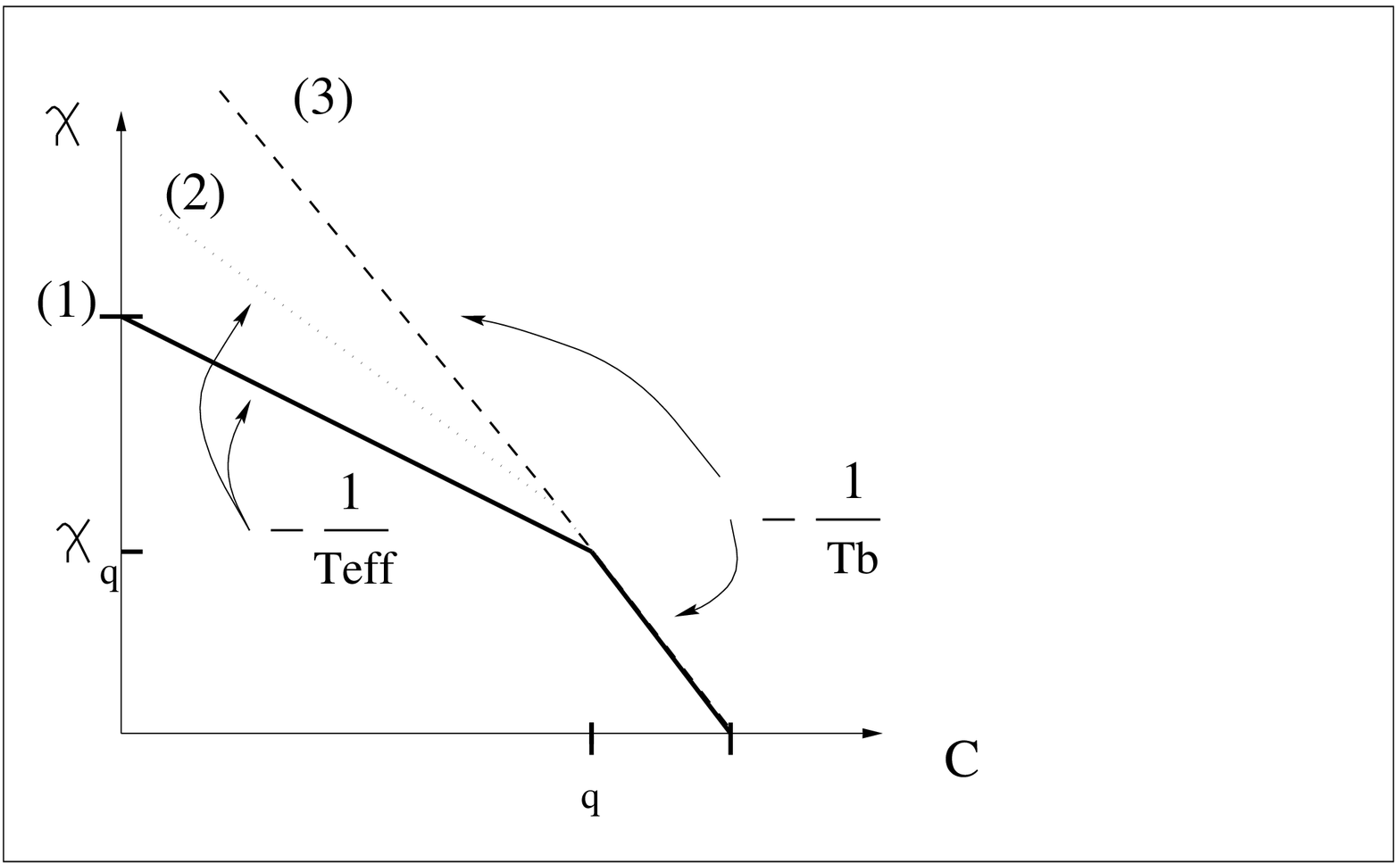}
}
\caption{Effective temperature plot.}
\label{coinse}
\end{figure}

The parametric response-versus correlation  plots like Fig.6 
reveal at a glance  the nature of the aging
dynamics. They were introduced  in the context of
analytically solved models of aging \cite{Cuku2} 
(see Section 3), and 
 were subsequently  
 numerically implemented for realistic models  \cite{checks}, for which 
an analytical solution is not available, with strikingly similar 
 results \cite{footteo}.

Let us argue that   $T_{eff}$ obtained from Fig. 6 is
 a  true temperature for the structural
motion,
in the sense that it has the good thermometry and heat exchange properties. 
It can be shown that \cite{Cukupe}:

\begin{itemize}

\item
 Any thermometer coupled to an observable in the system and tuned
to respond to the timescale of structural relaxation measures exactly
the $T_{eff}$ associated with its fluctuations and response. 

\item A frequency filter  tuned
 to the timescale of structural relaxation
 coupled to an observable in the system
and to an auxiliary  thermal bath of temperature $T^*$
will transfer energy from the highest to the lowest of
 $(T_{eff},T^*)$.

\end{itemize}

 These two results are model-independent. In addition,
within the framework of section 3 it has been shown that:

\begin{itemize}

\item Every pair of  observables in the system 
gives the same fluctuation-dissipation temperatures $T_{eff}$
at a given timescale.

\end{itemize}

These three facts characterise the FDT-related effective temperature
as {\em the only} bona fide temperatures associated to the slow
motion.
Moreover, there is a lesson we can learn from
 exactly  solvable models: $T_{eff}$ turns out to be
 of the order of, {\em but not
 quite equal to} the temperature at which the system falls out of 
equilibrium. 
We have here an instance of a fact mentioned at the beginning of the
article: the old fictive temperature idea is promoted to a
directly testable concept with thermodynamic (zero-th law) properties,
but is also somewhat modified.

\vspace{1cm}

{\bf 2. Ensembles and complexity}

\vspace{1cm}

Let us discuss now what are the suggestions of Fig. 6 from the point of
view of phase-space distributions.
We can picture a two-timescale dynamics as a quasi-equilibrium motion
 for the  fast evolution `inside the instantaneous cages' 
 (corresponding to the interval of
correlations $C>q$) combined with a  slow {\em structural} 
motion of the configuration of the cages.
The moment the system falls out of equilibrium,  the latter
 is not able to keep track ergodically.
It is  thus perfectly general that we should obtain a good 
equilibrium   form in Fig. 6 for the motion in the interval $C>q$,
 and  not for $C<q$.
 
What is not general, and begs for an explanation, is that the
structural, `cage rearrangement' motion behaves, 
from the point of view of FDT,
 as {\em thermalised to another temperature}, itself possibly decreasing 
adiabatically.
This immediately raises the question of whether there is a new, hidden
 form of
`ergodicity' for the structural motion, allowing us to construct a
statistical mechanics ensemble for the slow motion at each
energy level. 
 
 One such  construction is better introduced in the context in which it was
originally formulated by Edwards \cite{Sam}.
 Consider a granular medium that is compactifying under gentle
`tapping' \cite{Nagel}. After each tap, the particles settle into a blocked
configuration.

 Now, Edwards proposed
the extremely strong hypothesis  that the blocked configurations
reached dynamically at each density are the {\em typical} ones:
at any given time,
all macroscopic observables can be obtained by averaging their value
over all blocked configurations of the appropiate density.
This is an implicit definition of an ensemble for the dynamics. 
Once  accepted, one immediately is led to defining  an entropy
(in the glass language a `configurational entropy'), as the logarithm
of the number of blocked configurations of given volume --- and  energy, in
the case of soft particles.
Differentiating the configurational entropy with respect to the volume 
we obtain a `compactivity', and with respect to energy an inverse
{\em configurational temperature}.
  
This example of a  granular medium is particularly simple because 
by looking at the system in repose some time after each tap, 
 one concentrates on the slow rearrangements 
without having to deal with a  superposition of simultaneous  
fast and slow motion.

 Within the mean-field/mode-coupling scenario  to be discussed
 in Sect. 3, it is possible to calculate averages over all the energy minima
(the relevant blocked configurations for a zero temperature dynamics),
 as well as the their number at each energy level.
 The remarkable result is that the
configurational temperature thus calculated \`a la Edwards and 
 the  FDT temperature $T_{eff}$ obtained after a deep quench 
 are found to {\em coincide exactly}.
 Furthermore, the out of equilibrium macroscopic
 observables are correctly given
 by flat averages over the blocked regions 
\cite{Remi,stabarbara,Theo,Frvi,felix}.

 These models teach  us something else: for quenches to non-zero temperature,
it is not the energy minima that have to be computed in order to reproduce
the dynamic result, but rather the number of metastable states.
 This immediately poses a problem:
 while in the mean-field models a metastable state is unambiguously
defined as a region in phase space from which the system never
escapes,  for realistic, finite-dimensional
models at $T>0$ all metastable states have a finite lifetime.
In other words, there is no absolute separation between cage and
structural motion and
there is an {\em essential} ambiguity in the concept of metastable
state.
 This ambiguity carries
over to the concept of configurational entropy, and hence to the
configurational temperature, as the number of
states we count depends on the minimal lifetime we expect from a state
to call it a `state' \cite{Frvi,Biku}.

 This is a rather rapidly developing and not fully settled field, 
 several alternatives have been suggested but many questions remain open.
 One strategy that is easy to implement numerically 
is to identify as a state the Gibbs-Boltzmann distribution restricted 
to the set of configurations in the basin of attraction of
each energy minimum --- the `inherent structure'
 construction \cite{inherent_equilibrium}. 
 Though this identification is not without 
 problems of principle \cite{enfants}, 
the results in connection with the
aging dynamics are encouraging \cite{inherent_aging,felix}. 
Recently, a more direct   comparison   between
 compaction results and Edwards ensembles  in a schematic 
three dimensional model has
 yielded very good agreement \cite{Bakulose}.

Another strategy  is to use a direct 
definition of `states with a given lifetime' \cite{Frvi,Biku}.
Using a construction of
 Gaveau and Schulman \cite{larry}, one can compute averages over
states having lifetimes of a any preassigned value 
\cite{Biku}.  
Yet another possibility is to directly construct a phenomenological 
two-temperature thermodynamics, without invoking a statistical
mechanical ensemble \cite{Theo}.

Let us conclude by remarking that  the role played by  configurational
entropy here is stronger than in the 
 Kauzmann picture of an ideal
glass transition, since it  involves 
 the out of equilibrium dynamics, and not only  an infinitely slow cooling. 
This has an advantage with respect to robustness, since the present
scenario would be relevant in situations in which a thermodynamic glass
transition  either does not exist or is unobservable at
experimentally
accesible times.

At any rate, let us insist again, these ideas of generalised ensembles
can only be taken seriously  to the extent that the existence  of  FDT
temperatures, independent of the observable at each timescale,
is verified experimentally.

\vspace{1cm}

{\bf 3. Methods and Models}

\vspace{1cm}

Within a set of approximations  and  models the scenario
discussed above holds rigorously.
While these approximations are far from perfect, they have suggested 
tests that have been performed on realistic systems, with a
considerable degree of success.

The paradigmatic and best studied model of supercooled liquids with
two-step relaxations is the mode-coupling theory (MCT) \cite{mocu}.
These are  approximate  equations for the two-point
dynamical correlation functions, which can be obtained
through addition of a partial (though infinite) set of terms
in the perturbative expansion  of the exact dynamics \cite{review}.

An apparently different point of view was developed by Kauzmann, and Adam,
Gibbs and DiMarzio \cite{Kauzmann}, who first pictured
 the total entropy as composed of `rapid' and
`configurational' degrees of freedom, and the `ideal glass' 
transition as the temperature
at which the system runs out of configurational entropy. 

In a remarkably insightful  set of papers, Kirkpatrick,
 Thirumalai and Wolynes \cite{KTW}
showed that both approaches can be seen as aspects of the same theory.
To do this, they first noted that MCT is the {\em exact} solution of
the two-time  correlations of the high temperature  phase of 
a family of fully connected models having spin glass-like
 Hamiltonians but with multi-spin interactions.
(This is why we referred above to the {\em `mean-field/mode-coupling'}
 models).
They then showed that
 the partition function of these models ignores completely the
dynamic
transition at   $T_{MCT}$, but has an equilibrium transition at a
lower temperature $T_K$.
They explained this fact by showing that between  $T_{MCT}$ and $T_K$
the phase-space is composed of exponentially many,  mutually
inaccesible states  that the Gibbs measure confuses into a single, large,
deceivingly liquid-like state.
The equilibrium transition at $T_K$ happens because 
at just that temperature the number of states contributing to the
measure becomes of order one: precisely Kauzmann's  idea with the
logarithm of the number of states playing the role of configurational 
entropy. The paradigm of this form of transition is Derrida's Random
Energy Model \cite{Mepavi}. 

A natural question that arises immediately is, what does a system
satisfying exactly mode coupling equations for $T>T_{MCT}$ do  
 if quenched below the dynamic glass transition 
 $T<T_{MCT}$? The answer is  that
it {\em ages}\cite{Cuku}: the correlation functions depend on the waiting
time forever. Furthermore, the FDT plot of Fig. 6 converges to
two a strict two-temperature behaviour, with a $T_{eff}$ of the order
of (but not equal to) the transition temperature.

 As mentioned before,
 one can study the free energy minima \cite{criso} 
of  the corresponding model: flat
averages over metastable  states of the appropiate energy 
give the correct out of equilibrium values 
of macroscopic observables
\cite{Remi,stabarbara,Theo,Frvi}. Furthermore,
  the  configurational temperature obtained 
as the logarithmic derivative of the number of states with
respect to the free energy turns out to  give  exactly the dynamically
 defined $1/T_{eff}$.
Hence, within these models 
 Edwards' `compactivity' ideas are also strictly realised.

 Now, it is a  well known fact that the mode coupling transition 
 becomes a crossover  
 in real life \cite{mocu}. This  also has  an explanation  when the nature
 of the models is considered. The transition at $T_{MCT}$ relies on the 
 existence of fully stable excited
 states, but these will eventually decay  in any finite-dimensional 
 model through  nucleation \cite{KTW}.
 On the other hand, while the
  equilibrium transition at $T_K$ may or may not  in a realistic
model become a crossover as well, there is no {\em a priori} reason for this
 to be the case.

There are several strategies starting from  these considerations.
Given the limitations of the models considered, {\em it is first mandatory 
 to check directly with a simulation of a realistic
model} (say, a Lennard-Jones liquid) 
 whether a two-temperature aging scenario
remains valid in the presence of activated processes that may
eventually lead the system to thermalisation.  The evidence seems to
be positive \cite{checks}.
A much harder numerical check concerns the nature of the equilibrium
phase at low temperatures \cite{landscape}, using such small
samples as one can thermalise.

From the analytic point of view, there are various different 
 approaches being pursued.
Perhaps the most basic is to try to understand
 whether there is a generic reason that there should be 
only  one FDT temperature for all observables per widely separated
 timescale --- 
 independently from the
approximation scheme in which this scenario  was first 
(perhaps accidentally)  found. 

On a more practical line,
there are several schemes for improving the approximations for the
dynamics by considering diagrammatic ressumations \cite{review},
 although dealing with
 activated processes is notoriously hard.
One can also construct better approximations for the equilibrium
 calculations. This approach, 
  advocated by Mezard and Parisi 
\cite{MP}, may be motivated by the following consideration:
Clearly,   an equilibrium  calculation
is relevant at such late times   that energy, volume, and all macroscopic
observables
 can be considered to have relaxed to their thermodynamic equilibrium value.
If a glassy system has a transition below which $t_\alpha$ diverges
 strictly, even if all quantities have reached essentially 
their equilibrium value their $T_{eff}$ might still never reach the
bath's temperature:
 It turns out that in such cases we can calculate the limiting
 $t \rightarrow \infty$ form of Fig. 6 using equilibrium  techniques
\cite{silvioth},
and the result is directly related to Parisi's order parameter.
(Obviously, within this approach, `strong' glasses having
 $T_o=0$ are described as liquids).
This rather surprising connection between equilibrium and the last
stages of  out of equilibrium relaxation involves certain
general  assumptions, and in the specific case of
structural glasses it requires the  exclusion 
  from the equilibrium calculation of all configurations having a
degree of crystallisation.

\vspace{1cm}

{\bf 4. Experiments}

\vspace{1cm}

As remarked in several places above, the main observable
feature of the present scenario for glasses
is the existence of well separated timescales and a {\em
  single} fluctuation-dissipation temperature per timescale,
shared by all observables.
For example, 
the relation between spontaneous diffusion and mobility in a given
time interval should be the same for different kinds of particles
within the glass, or between rotational and translational degrees of
freedom of the same particle.
This relation defines a temperature that only coincides 
with the bath's temperature for the fastest relaxations.

The fluctuation-dissipation
experiments we have so far \cite{exper} are comparisons of
the dielectric susceptibility and polarisation noise in
low-temperature glycerol and in Laponite gel, respectively.
They show clearly that the low frequency ratio between noise amplitude
and susceptibility
 stays different from the  bath temperature after very long
times. The analysis of the data is somewhat complicated by the fact
that, for experimental reasons, these  results are given in the
 frequency domain, and it is difficult to unravel the fast and slow 
timescales.

Ideally, one should go on to
measure the time correlation between several fluctuating
quantities and compare these with the associated responses --- and
check that the proportionality factor is the same 
at equal timescales for all observables.
{\em If, and when, this is not the case, the  scenario discussed in this
paper does not apply}.
  
The measurements are always complicated by the waiting-time
dependence of correlations and responses. An approach that may be
useful in practice is to study  glasses that are gently driven,
 for example by shearing, and to use the surprising
 fact  that in
 such conditions aging can be stopped  
 without changing the response versus correlation curve \cite{stabarbara}.


\end{document}